\documentclass[aps,prc,groupedaddress,showpacs,preprint]{revtex4}
\usepackage{graphicx,bm}

\begin{document}

\title{On a redundancy in the parity-violating 2-nucleon contact Lagrangian}

\author{L. Girlanda}
\email[]{girlanda@pi.infn.it}
\affiliation{INFN, Sez.~di Pisa, Largo Bruno Pontecorvo, 56127 Pisa, Italy}

\date{\today}

\begin{abstract}
It is shown that the number of independent operators in the
nucleon-nucleon parity violating contact Lagrangian at the leading
order in the low-momentum expansion, can be reduced to five, by
using Fierz rearrangements of the nucleon fields.
\end{abstract}


\maketitle

Recent interest in the study of hadronic parity violation has been
triggered by the planning of new experiments involving few-body
systems, like NPDGamma at LANSCE \cite{npdgamma} or neutron spin rotation on $^4$He at
NIST \cite{nist}, in conjunction with the enormous
progress in the theoretical control on ab-initio calculations on such 
systems \cite{fewbody}. The traditional theoretical
framework for such studies was set by Desplanques, Donoghue and
Holstein in Ref.~\cite{ddh}, in the form of
a meson exchange model, generally known as the DDH model. More
recently an effective field theory description of nuclear parity
violation has been proposed in Ref.~\cite{holstein,musolf}, in which
nucleons and pions interact according to the dictates of chiral
symmetry (see Ref.~\cite{chptreview} for a comprehensive review). At
sufficiently low energies, pions can be integrated out 
and their contribution (as well as the ones from higher-mass
particles) is subsumed in the values of the coupling constants of the
contact interactions among nucleons (``pionless effective field
theory''). In either case the Lagrangian containing the nucleon
contact interactions appears as a crucial ingredient in the effective
field theory. 

In the heavy baryon formalism the most general
Lagrangian yields redundant operators, which have to be eliminated by
imposing reparameterization invariance \cite{luke}. Alternatively, one
can start from the relativistic theory and perform the
non-relativistic reduction afterwards, as was done in
Ref.~\cite{holstein}.

The two-nucleon contact interactions can be classified according to
their transformation properties under the chiral symmetry
\cite{kaplan}. The possible flavor structures that can arise change
the isospin by $\Delta I=0,1,2$:
\begin{equation}
\begin{array}{lll}
\Delta I=0:  \quad 
\begin{array}{l}
F_1^{ij,kl} = \delta_{ij} \delta_{kl},  \\
F_3^{ij,kl} = \tau^a_{ij} \tau^a_{kl} , 
\end{array} &
\Delta I=1:  \quad 
\begin{array}{l}
F_{2'}^{ij,kl} = \delta_{ij} \tau^3_{kl} ,\\
F_{4'}^{ij,kl} = \tau^3_{ij} \delta_{kl} , \\
F_6^{ij,kl} = i \epsilon^{ab3} \tau^a_{ij} \tau^b_{kl} ,
\end{array} &
\Delta I=2:  \quad 
F_5^{ij,kl} = {\cal I}^{ab} \tau^a_{ij} \tau^b_{kl},
\end{array} 
\end{equation}
where $i,j,k,l$ are the isospin indices of the four nucleons and 
the matrix ${\cal I}={\mathrm{diag}}(1,1,-2)$.
Notice that all these structures are symmetric under the exchange of
$i\leftrightarrow j$ and $k\leftrightarrow l$, except $F_6^{ij,kl}$,
which is antisymmetric.
It is also convenient to classify the flavor structures according to
the symmetry properties under the exchange of
$(ij)\leftrightarrow(kl)$ and use, instead of $F_{2'}^{ij,kl}$ and
$F_{4'}^{ij,kl}$, the structures
\begin{equation}
F_{2}^{ij,kl}=\delta_{ij}\tau^3_{kl}+\delta_{kl}\tau^3_{ij}, \quad
F_{4}^{ij,kl}=\delta_{ij}\tau^3_{kl} - \delta_{kl}\tau^3_{ij},
\end{equation}
so that 
\begin{equation}
F_i^{ij,kl} = s_i F_i^{kl,ij},\quad s_1=s_{2}=s_3=-s_{4}=s_5=-s_6=1.
\end{equation}

One can easily list all possible Lorentz invariant four-nucleon
operators which violate P and conserve CP. Those containing up to one
space-time derivative are listed below ($i=1,...,5$ henceforth),
\begin{equation}
\begin{array}{ll}
O_i^{(1)} = F_i \otimes \bar \psi \gamma^\mu \psi \bar \psi \gamma_\mu
\gamma_5 \psi, & O_6^{(2)} = F_6 \otimes \bar \psi \psi \bar \psi \gamma_5 \psi
,  \\
\tilde O_i^{(1)} = F_i \otimes \bar \psi
i \stackrel{\leftrightarrow}{\partial}_\mu \psi \bar \psi \gamma^\mu
\gamma_5 \psi, & \tilde O_6^{(5)} = F_6 \otimes \bar \psi \gamma_5
i \stackrel{\leftrightarrow}{\partial}_\mu \psi \bar \psi \gamma^\mu
\psi,  \\
\tilde O_i^{(2)} = F_i \otimes \bar \psi
\gamma^\mu \gamma_5 \psi \partial^\nu ( \bar \psi \sigma_{\mu\nu}
\psi) , & \tilde O_6^{(6)} = F_6 \otimes \bar \psi
\gamma^\mu  \psi \partial^\nu ( \bar \psi \sigma_{\mu\nu} \gamma_5
\psi) ,  \\
\tilde O_i^{(3)} = F_i \otimes \bar \psi \gamma_\mu 
\stackrel{\leftrightarrow}{\partial}_\nu \psi \bar \psi
\sigma^{\mu\nu} \gamma_5 \psi, & \tilde O_6^{(7)} = F_6 \otimes \bar \psi \gamma_\mu \gamma_5 
\stackrel{\leftrightarrow}{\partial}_\nu \psi \bar \psi
\sigma^{\mu\nu}  \psi,  \\
\tilde O_i^{(4)} = F_i \otimes \bar \psi \gamma_\mu 
 \psi \bar \psi
 \stackrel{\leftrightarrow}{\partial}_\nu\sigma^{\mu\nu} \gamma_5
 \psi, & \tilde O_6^{(8)} = F_6 \otimes \bar \psi \gamma_\mu \gamma_5 
 \psi \bar \psi \stackrel{\leftrightarrow}{\partial}_\nu
\sigma^{\mu\nu} \psi.
\end{array}
\end{equation}
Some of these operators are related by the field equations of
motion. Using the field equations of motion to reduce the number of
independent operators corresponds to the freedom of redefining the
interpolating fields of the effective theory. One can establish the
following relations,
\begin{eqnarray}
\tilde O_i^{(2)} &=& s_i \left[ 2 m O_i^{(1)} - \tilde O_i^{(1)}\right],
\\
\tilde O_i^{(3)} &=& -\tilde O_i^{(2)}, \\
\tilde O_i^{(4)} &=& - 2 m O_i^{(1)},  \\ 
\tilde O_6^{(5)} &=& \tilde O_6^{(6)}, \\ 
\tilde O_6^{(7)} &=&  - \tilde O_6^{(6)}, \\
\tilde O_6^{(8)} &=& 2 m O_6^{(2)}, 
\end{eqnarray}
so that everything can be expressed as combinations of the 12 operators
$O_{1,...,5}^{(1)}$, $O^{(2)}_6$, $\tilde O^{(2)}_{1,...,5}$ and
$\tilde O^{(6)}_6$, which correspond to the set identified in
Appendix~A of Ref.~\cite{holstein}. Based on the above set of
operators, a non-relativistic  Lagrangian  containing 11 operators was
introduced (cfr. Eq.~(71) of Ref.~\cite{holstein}), and used to derive
a potential containing 10 operators (cfr. Eq.~(5) of
Ref.~\cite{holstein} or Eq.~(9) of Ref.~\cite{musolf}).

This set of operators is complete but redundant, as can be shown
by using Fierz rearrangements. Let us start with the flavor
structures. By interchanging
the indices $j$ and $l$, the following relations hold:
\begin{eqnarray}
2 F_1^{il,kj} &=&  F_1^{ij,kl} +  F_3^{ij,kl} , \\
 F_{2}^{il,kj}&=&  F_{2}^{ij,kl} , \\
 F_{4}^{il,kj}&=&   F_6^{ij,kl}, \\
F_5^{il,kj} &=& F_5^{ij,kl}.
\end{eqnarray}
 
For the spinor structures of interest, the relevant Fierz identities
\cite{fierz} are conveniently written as
\begin{eqnarray}
() [\gamma_5] &=&
\frac{1}{4}  \biggl\{ 
(\gamma_5] [ ) + (][\gamma_5) + ( \gamma^\mu \gamma_5] [\gamma_\mu) -
    (\gamma^\mu] [ \gamma_\mu \gamma_5) +\frac{1}{2}
    (\sigma^{\mu\nu}][\sigma_{\mu\nu}\gamma_5) 
\biggr\}  \\
(\gamma^\mu) [ \gamma_\mu \gamma_5] &=&
(\gamma_5] [ ) - (][\gamma_5) -\frac{1}{2} ( \gamma^\mu \gamma_5]
  [\gamma_\mu) - \frac{1}{2}    (\gamma^\mu] [ \gamma_\mu \gamma_5)
 \\
(\sigma_{\mu \nu} ) [\gamma^\mu \gamma_5 ] &=&
\frac{3}{4} i \biggl\{ 
- (  \gamma_\nu \gamma_5][) 
+ ( \gamma_\nu][\gamma_5)
+ (\gamma_5 ] [ \gamma_\nu) 
+ ( ] [ \gamma_\nu \gamma_5) 
\biggr\} \nonumber \\
&& + \frac{1}{4}\biggl\{ 
-  (\sigma_{\mu \nu} ] [  \gamma^\mu \gamma_5) 
- (  \gamma^\mu \gamma_5][\sigma_{\mu\nu}) 
- (\gamma_5  \sigma_{\mu\nu}] [ \gamma^\mu) 
+ ( \gamma^\mu][\gamma_5 \sigma_{\mu\nu})
\biggr\},  \\
 () [  \gamma_\nu \gamma_5 ] &=& 
\frac{1}{4}  \biggl\{ 
(  \gamma_\nu \gamma_5][) 
- ( \gamma_\nu][\gamma_5) 
+ (\gamma_5 ] [ \gamma_\nu) 
+ ( ] [  \gamma_\nu \gamma_5) 
\biggr\} \nonumber \\ 
&& + \frac{i}{4}\biggl\{ 
- (\sigma_{\mu \nu} ] [  \gamma^\mu \gamma_5) 
+ (  \gamma^\mu \gamma_5][\sigma_{\mu \nu}) 
- (\gamma_5  \sigma_{\mu \nu}] [ \gamma^\mu) 
- ( \gamma^\mu][\gamma_5 \sigma_{\mu \nu}) 
\biggr\},   \\
 (\gamma^\mu) [ \gamma_5 \sigma_{\mu \nu} ] &=& 
\frac{3}{4} i \biggl\{ 
( \gamma_\nu \gamma_5  ][) 
+ ( \gamma_\nu][\gamma_5)
- (\gamma_5 ] [ \gamma_\nu) 
+ ( ] [  \gamma_\nu \gamma_5) 
\biggr\} \nonumber \\
&& + \frac{1}{4}\biggl\{ 
 (\sigma_{\mu \nu} ] [  \gamma^\mu \gamma_5) 
- (  \gamma^\mu \gamma_5][\sigma_{\mu \nu}) 
- (\gamma_5  \sigma_{\mu \nu}] [ \gamma^\mu) 
- ( \gamma^\mu][\gamma_5 \sigma_{\mu \nu})
\biggr\},   \\
 (\gamma_5) [ \gamma_\nu ] &=& 
\frac{1}{4}  \biggl\{ 
-(  \gamma_\nu \gamma_5][) 
+ ( \gamma_\nu][\gamma_5)
+ (\gamma_5 ] [ \gamma_\nu) 
+ ( ] [  \gamma_\nu \gamma_5) 
\biggr\} \nonumber \\
&& + \frac{i}{4}\biggl\{ 
- (\sigma_{\mu \nu} ] [  \gamma^\mu \gamma_5) 
- (  \gamma^\mu \gamma_5][\sigma_{\mu \nu}) 
- (\gamma_5  \sigma_{\mu \nu}] [ \gamma^\mu) 
+ ( \gamma^\mu][\gamma_5 \sigma_{\mu \nu})
\biggr\},  
%
%
\end{eqnarray}
where ``$($'', ``$)$'', ``$[$'' and ``$]$'' are shorthands for $\bar
\psi_1$, $\psi_2$, $\bar \psi_3$ and $\psi_4$ respectively and 
 an overall minus sign should be included due do the
anticommuting nature of the fermion fields.
Combining the flavor and spinor indices rearrangements, we have the
operator identities
\begin{eqnarray}
 O^{(1)}_1 &=& O^{(1)}_3,  \\
 O^{(1)}_{4'} &=&  2 O^{(2)}_6,  \\
4 m O^{(1)}_1 &=& 3 \tilde O^{(2)}_1 + \tilde O^{(2)}_3,  \\
\tilde O^{(2)}_{2} &=& m O^{(1)}_{2},  \\
\tilde O^{(2)}_{5} &=& m O^{(1)}_{5},  \\
\tilde O^{(2)}_{4} &=& - 2 m O^{(2)}_{6} - \tilde O^{(6)}_6.
\end{eqnarray}
Thus the number of independent operators is reduced to 6.
 
 In the leading order of the non-relativistic reduction, the following
relations hold
\begin{eqnarray}
 O_i^{(1)} &=& \frac{1}{2 m} \left[ - N_i^{(1)} + s_i N_i^{(2)} - s_i N_i^{(3)}
   \right],  \\
 O_6^{(2)} &=& \frac{1}{2 m} N_6^{(1)},  \\
 \tilde O_i^{(1)} &=&- N_i^{(1)} + s_i N_i^{(2)},  \\
 \tilde O_i^{(2)} &=& - N_i^{(3)},   \\
 \tilde O_i^{(3)} &=& s_i N_i^{(3)},  \\
 \tilde O_i^{(4)} &=& N_i ^{(1)} - s_i N_i^{(2)} + s_i N_i^{(3)},  \\
 \tilde O_6^{(5)} &=& -N_6^{(1)},  \\
 \tilde O_6^{(6)} &=& -N_6^{(1)},  \\
 \tilde O_6^{(7)} &=& N_6^{(1)},  \\
 \tilde O_6^{(8)} &=& N_6^{(1)},  
\end{eqnarray}
with the non-relativistic operators defined by
\begin{equation}
\begin{array}{l}
N_i^{(1)} =F_i \otimes N^\dagger N N^\dagger i \stackrel{\leftrightarrow}{\nabla}\cdot
\vec{\sigma} N,   \\
N_i^{(2)} =F_i \otimes N^\dagger \vec{\sigma} N \cdot N^\dagger i
\stackrel{\leftrightarrow}{\nabla} N, \\
N_i^{(3)} = F_i \otimes \epsilon_{ijk} N^\dagger \sigma^i N \nabla^j
( N^\dagger \sigma^k  N), \\
N_6^{(1)} = F_6 \otimes i \vec{\nabla} ( N^\dagger N) N^\dagger
\vec{\sigma} N.
\end{array}
\end{equation}
One can then make use of the fact that the operators $O_6^{(2)}$ and
$\tilde O_6^{(6)}$ give rise to the same non-relativistic structure, 
as already observed in Ref.~\cite{holstein}. Therefore, the number of
independent operators  can be further  reduced to 5 up to
order $O(Q)$, and  the minimal PV two-nucleon non-relativistic contact
Lagrangian may be taken to assume the form
\begin{eqnarray}
{\cal L}_{PV,NN} &=& \frac{1}{\Lambda_\chi^3} \left\{ C_1 ( N^\dagger
\vec{\sigma} N \cdot N^\dagger i \stackrel{\leftrightarrow}{\nabla} N
- N^\dagger
 N  N^\dagger i \stackrel{\leftrightarrow}{\nabla} \cdot \vec{\sigma} N
 ) \nonumber \right.\\
&& - \tilde C_1 \epsilon_{ijk} N^\dagger \sigma^i N \nabla^j
( N^\dagger \sigma^k  N) \nonumber \\
&&- C_{2} \epsilon_{ijk} [N^\dagger \tau_3 \sigma^i N \nabla^j
( N^\dagger \sigma^k  N) + N^\dagger  \sigma^i N \nabla^j
( N^\dagger \tau_3 \sigma^k  N)] \nonumber \\
&& 
- \tilde C_5 {\cal I}_{ab} \epsilon_{ijk} N^\dagger \tau^a \sigma^i N \nabla^j
( N^\dagger \tau^b \sigma^k  N) \nonumber \\
&& \left.
+C_6 \epsilon^{ab3} \vec{\nabla} (N^\dagger \tau^a N )\cdot  N^\dagger \tau^b
\vec{\sigma} N  \right\},
\end{eqnarray}
where the notations for the coupling constants have been chosen so as
to conform to Ref.~\cite{holstein} except for $C_{2}$ which replaces
$C_2+C_4$. 
Thus, for instance, in the notations of Ref.~\cite{holstein}, the term 
\begin{equation}
N^\dagger \tau^a N N^\dagger \tau^a \vec{\sigma} \cdot i \vec{D}_- N -
N^\dagger \tau^a i \vec{D}_- N \cdot N^\dagger \tau^a \vec{\sigma}
N, \quad [ N^\dagger i D^\mu_\pm N \equiv ( i D^\mu N )^\dagger N \mp
  N^\dagger ( i D^\mu N)],
\end{equation}
in the non-relativistic reduction of the Lagrangian can be omitted, in
as much the same way as a term
\begin{equation}
N^\dagger \tau^a \vec{\sigma} N \cdot N^\dagger \tau^a \vec{\sigma}  N
\end{equation}
can be omitted in the parity-conserving sector \cite{vankolk}.

It is worth noting that the reduction of the number of independent
operators down to five has no practical consequences
at the present stage of phenomenological analyses since
only five combinations of low-energy constants are relevant at low
energies. This was already noticed in Refs.~\cite{holstein,musolf}
 on the basis of the
observation that, since only S- and P-wave amplitudes are important in this
regime, several operators give rise to identical matrix elements (see
also the discussion in Ref.~\cite{liu}). Nevertheless, in view of a
description of nuclear parity violation with the chiral effective theory,
it is important to use a truly minimal set of operators, and have a
one-to-one correspondence between physical observables and low-energy
constants of the effective Lagrangian. 

I thank Barry Holstein for correspondence, and Rocco Schiavilla and 
Michele Viviani for useful discussions.

\end{document}